\documentclass[conference]{IEEEtran}
\IEEEoverridecommandlockouts

\usepackage{cite}
\usepackage{amsmath,amssymb,amsfonts}
\usepackage{algorithmic}
\usepackage{graphicx}
\usepackage{textcomp}
\usepackage{xcolor}
\usepackage{url}
\usepackage{calc}
\usepackage{subcaption}
\usepackage[font=small,labelfont=bf]{caption}

\makeatletter
\newcommand*{\inlineequation}[2][]{%
  \begingroup
    \refstepcounter{equation}%
    \ifx\\#1\\%
    \else
      \label{#1}%
    \fi
    \relpenalty=10000 %
    \binoppenalty=10000 %
    \ensuremath{%
      #2%
    }%
    ~\@eqnnum
  \endgroup
}
\makeatother
\usepackage{hyperref}

\def\BibTeX{{\rm B\kern-.05em{\sc i\kern-.025em b}\kern-.08em
    T\kern-.1667em\lower.7ex\hbox{E}\kern-.125emX}}

\begin{document}

\title{EMulator: Rapid Estimation of Complex-valued Electric Fields using a U-Net Architecture\\
\thanks{This work is supported by The Robert and Janice McNair Foundation.}
}


\author{
\IEEEauthorblockN{Fatima Ahsan\IEEEauthorrefmark{1}, Lorenzo Luzi\IEEEauthorrefmark{1}, Richard G. Barainuk\IEEEauthorrefmark{1}, Sameer A. Sheth\IEEEauthorrefmark{4}, Wayne Goodman\IEEEauthorrefmark{2}, Behnaam Aazhang\IEEEauthorrefmark{1}}\\
    \IEEEauthorblockA{\IEEEauthorrefmark{1}Department of Electrical and Computer Engineering, Rice University, Houston, TX, USA.\\  \IEEEauthorrefmark{4}Department of Neurosurgery, Baylor College of Medicine, Houston, TX, USA.\\
    \IEEEauthorrefmark{2}Menninger Department of Psychiatry and Behavioral Sciences, Baylor College of Medicine, Houston, TX, USA.} \\

Email: \{fatima.ahsan, lorenzo.luzi, richb, aaz\}@rice.edu, \{sameer.sheth, wayne.goodman\}@bcm.edu
}

\newcommand\ms{\text{ ms}}

\newcommand\vect[1]{\mathbf #1}
\newcommand{\va}{\vect{a}}  
\newcommand{\vb}{\vect{b}}
\newcommand{\vc}{\vect{c}}  
\newcommand{\vd}{\vect{d}}  
\newcommand{\ve}{\vect{e}}
\newcommand{\vf}{\vect{f}}  
\newcommand{\vg}{\vect{g}}  
\newcommand{\vh}{\vect{h}}
\newcommand{\vi}{\vect{i}}  
\newcommand{\vj}{\vect{j}}  
\newcommand{\vk}{\vect{k}}
\newcommand{\vl}{\vect{l}}  
\newcommand{\vm}{\vect{m}}  
\newcommand{\vn}{\vect{n}}
\newcommand{\vo}{\vect{o}}  
\newcommand{\vp}{\vect{p}}  
\newcommand{\vq}{\vect{q}}
\newcommand{\vr}{\vect{r}}  
\newcommand{\vs}{\vect{s}}  
\newcommand{\vt}{\vect{t}}
\newcommand{\vu}{\vect{u}}  
\newcommand{\vv}{\vect{v}}  
\newcommand{\vw}{\vect{w}}
\newcommand{\vx}{\vect{x}}  
\newcommand{\vy}{\vect{y}}  
\newcommand{\vz}{\vect{z}}

\newcommand{\mA}{\vect{A}}  
\newcommand{\mB}{\vect{B}}
\newcommand{\mC}{\vect{C}}
\newcommand{\mD}{\vect{D}}
\newcommand{\mE}{\vect{E}}
\newcommand{\mF}{\vect{F}}
\newcommand{\mG}{\vect{G}}
\newcommand{\mH}{\vect{H}}
\newcommand{\mI}{\vect{I}}
\newcommand{\mJ}{\vect{J}}
\newcommand{\mK}{\vect{K}}
\newcommand{\mL}{\vect{L}}
\newcommand{\mM}{\vect{M}}
\newcommand{\mN}{\vect{N}}
\newcommand{\mO}{\vect{O}}
\newcommand{\mP}{\vect{P}}
\newcommand{\mQ}{\vect{Q}}
\newcommand{\mR}{\vect{R}}
\newcommand{\mS}{\vect{S}}
\newcommand{\mT}{\vect{T}}
\newcommand{\mU}{\vect{U}}
\newcommand{\mV}{\vect{V}}
\newcommand{\mW}{\vect{W}}
\newcommand{\mX}{\vect{X}}
\newcommand{\mY}{\vect{Y}}
\newcommand{\mZ}{\vect{Z}}

\newcommand\milliamp{\text{ mA}}
\newcommand\killohertz{\text{ kHz}}
\newcommand\voltsperMeter{\text{ V/m}}
\newcommand\cm{\text{ cm}}
\newcommand\mm{\text{ mm}}
\newcommand\GHz{\text{ GHz}}
\newcommand\Hz{\text{ Hz}}
\newcommand\MHz{\text{ MHz}}

\maketitle

\begin{abstract}
A common factor across electromagnetic methodologies of brain stimulation is the optimization of essential dosimetry parameters, like amplitude, phase, and location of one or more transducers, which controls the stimulation strength and targeting precision. Since obtaining in-vivo measurements for the electric field distribution inside the biological tissue is challenging, physics-based simulators are used. However, these simulators are computationally expensive and time-consuming, making repeated calculations of electric fields for optimization purposes computationally prohibitive. To overcome this issue, we developed EMulator, a U-Net architecture-based regression model, for fast and robust complex electric field estimation. We trained EMulator using electric fields generated by 43 antennas placed around 14 segmented human brain models. Once trained, EMulator uses a segmented human brain model with an antenna location as an input and outputs the corresponding electric field. A representative result of our study is that, at $\textbf{1.5} \GHz$, on the validation dataset consisting of 6 subjects, we can estimate the electric field with the magnitude of complex correlation coefficient of $\textbf{0.978}$. Additionally, we could calculate the electric field with a mean time of $\textbf{4.4 \ms}$. On average, this is at least $\times \textbf{1200}$  faster than the time required by state-of-the-art physics-based simulator COMSOL. The significance of this work is that it shows the possibility of real-time calculation of the electric field from the segmented human head model  and antenna location, making it possible to optimize the amplitude, phase, and location of several different transducers with stochastic gradient descent since our model is almost everywhere differentiable.
\end{abstract}

\begin{IEEEkeywords}
 electromagnetic wave propagation, physics modeling,  U-Net, data-driven electric field estimation
\end{IEEEkeywords}


\section{Introduction}

FDA-approved physical means of brain stimulation, such as deep brain stimulation~\cite{deeb2016proceedings}, transcranial direct current stimulation (tDCS)~\cite{nitsche2008transcranial}, and transcranial magnetic stimulation (TMS)~\cite{paulus2008state}, are gaining importance because of their increased precision in targeting the affected brain areas compared to the traditional course of medication~\cite{DBSvsMedications,No_medicines}. Furthermore, many other novel methodologies, such as EMvelop stimulation~\cite{ahsan2022emvelop, fatima2020} and ultrasound stimulation~\cite{giammalva2021focused}, are also under investigation. \par

To ensure focal and intense brain stimulation, essential dosimetry parameters, like amplitude, phase, and the location of one or more transducers, needs to be optimized~\cite{Dmochowski_2011, domo_2017_foc, focPaper_Accessi}. Obtaining in-vivo measurements for the electric field distribution inside the brain tissue is  challenging  since that requires invasive surgery~\cite{para_invivo}. Hence, physics-based solvers, such as COMSOL Multiphysics~\cite{COMSOL_software} and SimNIBS~\cite{simnibs}, are used to calculate the electric field induced by the transducers inside the brain tissue. 
This requires precise physics-based  finite-element-modeling (FEM), which is computationally expensive, as the imported models need to be discretized and meshed, and partial differential equations governing the underlying physics are solved iteratively. Additionally, any change in the brain model or location of the transducers would require recomputing the electric fields. Running such a computationally expensive FEM software  is inefficient for algorithms that need to be executed repeatedly, such as  transducer location optimization or uncertainty analysis.

In order to overcome this issue, several works have tackled the problem of electric field estimation. For example,~\cite{Akimasa_2019} used deep neural networks to estimate the electric fields inside the brain tissue from magnetic resonance image (MRI) scans of 32 subjects and determined prediction accuracy for 5 new subjects. However, only a subset of the brain tissue (motor cortex) was used in the training. Hence, the trained model could only predict electric field in a specific brain region. In~\cite{Aberra2022_TMS_thresholds}, investigators used convolutional neural networks to rapidly calculate the threshold needed  due to TMS excitation. In~\cite{jia2022deeptdcs}, authors used data-driven techniques to find the currents induced inside the brain due to tDCS.  However, all  above works perform electric field estimation only in low frequency regime ($\leq 10\killohertz$ range). 
Additionally, they do not tackle the problem of full-wave electromagnetic wave estimation, which requires computation of both the amplitude and phase of the electric field (complex electric field). Hence, there is a need to develop  more generic frameworks, that do not make the quasi-static approximation~\cite{wang2024quasistatic}, and can be used in applications where high frequency electric fields are generated~\cite{quasi_static_JNE}, such as the ones being investigated for neuromodulation~\cite{ahsan2022emvelop} and conduction block~\cite{wang2024quasistatic}. 


 \begin{figure}[htb!]
\captionsetup[subfigure]{font=scriptsize,labelfont=scriptsize}
\centering
\begin{subfigure}[b]{0.5\textwidth}
   \includegraphics[width=0.95\linewidth]{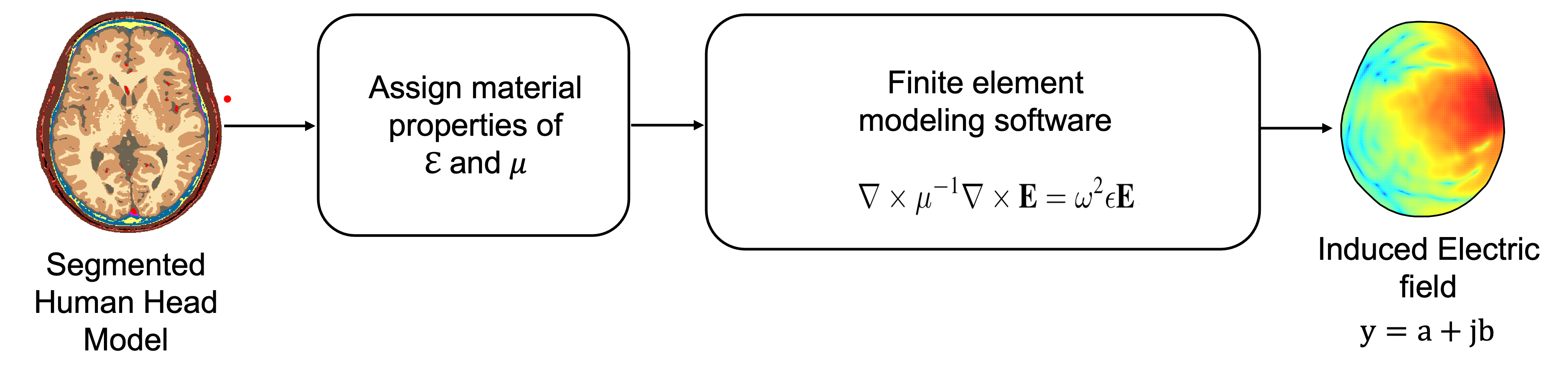}\vspace*{-2mm}
   \caption{Physics-based electric field estimation}
   \label{fig:phy} 
\end{subfigure}
\vspace*{2mm}
\begin{subfigure}[b]{0.5\textwidth}
   \includegraphics[width=0.95\linewidth]{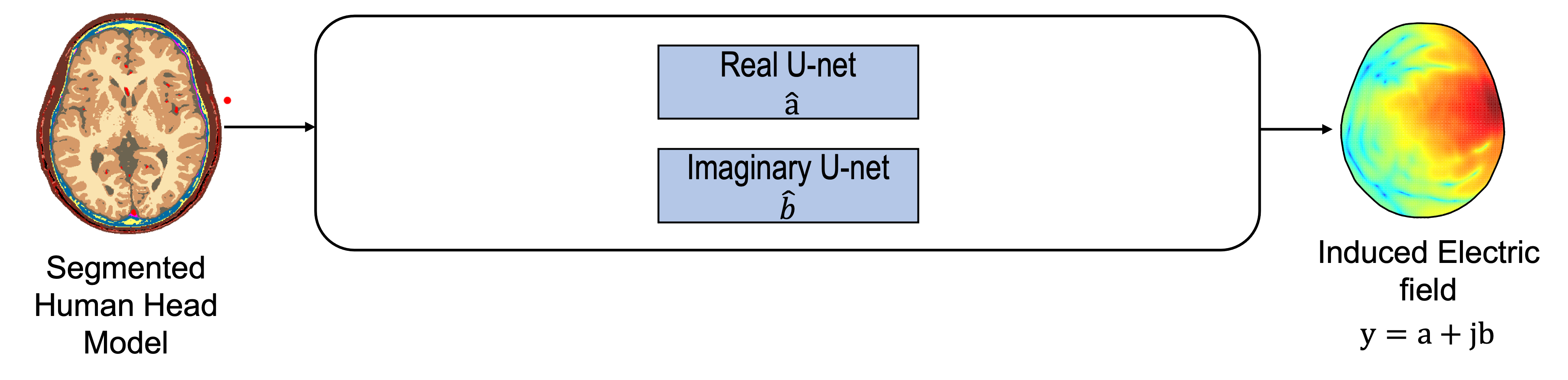}\vspace*{-4mm}
   \caption{Data-driven electric field estimation}
   \label{fig:datadd}
\end{subfigure}
\caption{Comparison of the physics-based and data-driven approaches.  The segmented human head model and the desired antenna location, shown as a red circle, are the inputs to the physics-based modeling software, which outputs the resultant complex electric field. After training, our data-driven regression model can estimate the electric field much faster than the physics-based software.}\label{fig:pipeline}
\end{figure}

Here, we investigate if we can calculate high frequency electric field inside segmented human brain models using data-driven models rather than running  computationally expensive physics-based solvers. This idea behind EMulator is conceptually shown in Fig.~\ref{fig:pipeline}.  To accomplish this, we import 20 segmented human brain models from BrainWeb~\cite{BrainWeb, BrainWebsite}. Then, we train our model EMulator,  a U-Net~\cite{ronneberger2015u} based architecture, using 14 of these subjects with antennas radiating at 43 specific locations around the brain tissue.
After training, when tested on the validation dataset consisting of 6 subjects, our model could estimate the electric field with the magnitude of complex correlation coefficient of $0.996, 0.989$. and $ 0.978$ at 400 MHz, 900 MHz, and 1.5 GHz, respectively. Additionally, we were able to estimate the electric field with a mean time of $4.3\ms$, $4.5\ms$, and $4.4\ms$ at $400\MHz$, $900\MHz$, and $1.5\GHz$, respectively. On average, this is at least $\times 1200$  faster than the computational time required by COMSOL. Additionally, our work  showed high correlation of $0.96$ and above for non-trained antenna locations as well.


Our results show that EMulator can serve as a fast and accurate tool to compute electric fields in the place of physics-based simulation software, such as COMSOL. Since EMulator is an almost everywhere differentiable model, it can used for a host of optimization procedures. We provide three useful potential applications of EMulator. First, EMulator can be used to optimize over the location of the antennas/transducers for brain stimulation methodologies. Second, this framework allows us to use stochastic gradient descent to optimize cost function for joint amplitude, phase, and location optimization. Finally, EMulator can be used for uncertainty analysis, as it allows us to rapidly generate electric fields from a set of coordinates for the antenna.

\section{Methods}\label{methods}
In this section, we explain the framework of the EMulator, our U-Net based model that directly learns a mapping from the given human head model and antenna location and outputs the corresponding electric field. Our goal is to learn the mapping $f$ using data-driven learning techniques and then use the trained model to predict the electric field $f(x_i)$ for a data point $x_i$ in the test set.

\subsection{Dataset generation}\label{EMu:datsetgen}

\begin{figure}[htb!]
\centering
\includegraphics[width=.13\textwidth, height=.15\textwidth]{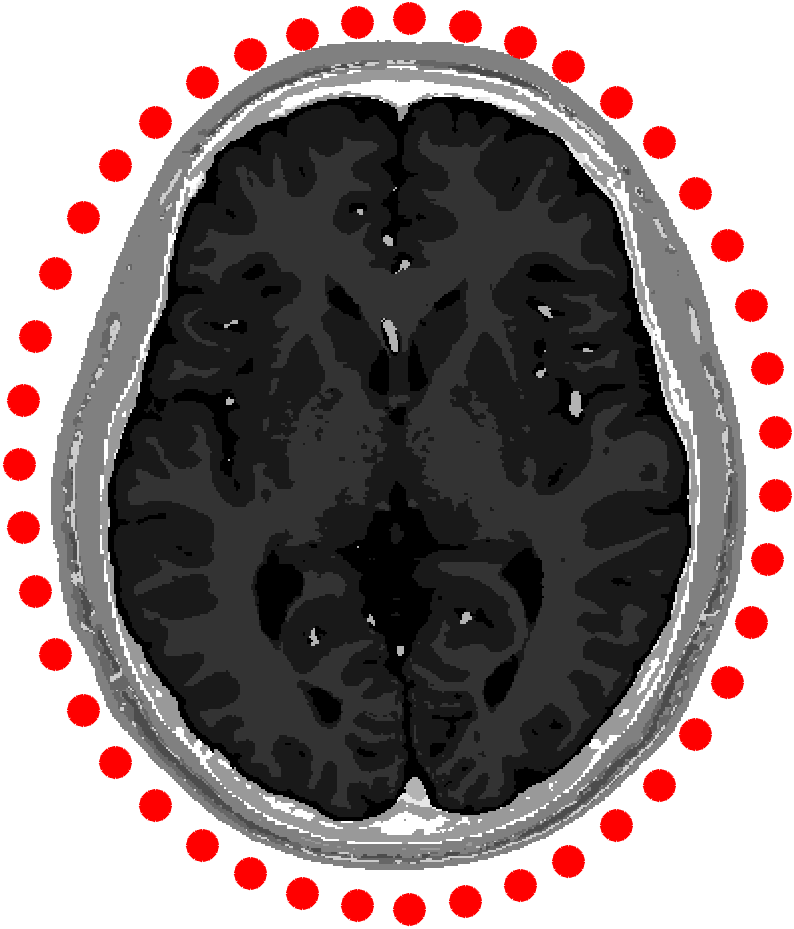}
\caption{Antennas over 43  locations around a representative human brain tissue plot are shown as red circles. For 14 training subjects, the electric field generated by these antennas is used in training EMulator. For 6 validation subjects, the electric field generated by these antennas is used in validating EMulator.}
\label{fig:43antzlocplot}
\end{figure}

We obtain MRIs of 20 subjects from BrainWeb~\cite{BrainWeb,BrainWebsite}. The MRI is segmented into twelve tissue types: CSF, grey matter, white matter, fat, muscle, muscle/skin, skull, vessels, connective tissue, dura mater, and bone marrow. 
Next, we import these segmented  human head models into physics-based FEM software COMSOL. For electromagnetic wave propagation in a medium, each material  is characterized by its permeability $\mu$ and complex permittivity $\epsilon$~\cite{EMwaveBook, Gabriel_1996}  at a particular frequency. We choose three frequencies that are reported in literature for use in biological tissues~\cite{ahsan2022emvelop, BrainInjur, Hyperthermia,  BreastCancer}: $400 \MHz$, $900 \MHz$ and $1.5 \GHz$. Hence, we assign the respective  $\mu$ and $\epsilon$ values to each of these tissues according to the IT'IS foundation database~\cite{itis}.  Since time-varying currents generate electromagnetic waves, we put dipole antennas at 43 locations around the brain tissue for generating EM waves, as shown in Fig.~\ref{fig:43antzlocplot}. When time-varying sinusoidal input currents at an angular frequency of $\omega$ are provided to these dipole antennas, the electric field  induced inside the brain tissue can be found by solving the  wave equation Maxwell's equation $\inlineequation[maxwellzeq]{ \nabla \times \mu^{-1} \nabla \times \mE = \omega^2 \epsilon  \mE }$, where $\mE$ denotes the calculated electric field. In order to compute $\mE$, COMSOL discretizes and solves (\ref{maxwellzeq}) on the entire segmented human brain model. The  electric field is collected with the spatial resolution of $1.5 \mm \times 1.5 \mm$ for a total of $158 \times 133$ voxels for each of the 43 antennas for all 20 subjects. This ground truth electric field is used to access the training and validation performance of our model.

\begin{figure*}[htb!]
\begin{minipage}{0.9\linewidth} 
\footnotesize \quad      \quad   \quad   \quad  \quad      \quad   \quad    \quad       \quad   \quad   \quad      $y$   \quad   \quad    \quad   \quad      \quad $\hat{y}$  \quad   \quad      \quad   \quad   \quad     \quad   \quad   AD \quad  \quad  \quad \quad  \quad   RD \quad  \quad  \quad \quad  \quad \quad  \quad    $\angle y$  \quad \quad  \quad \quad \quad  $\angle \hat{y}$ \quad  \quad  \quad \quad \quad  \quad \quad  PD
\end{minipage}
\begin{minipage}{\linewidth} 
\vspace{-10.0mm}
\scriptsize  \quad \quad  \quad  \quad \quad  \quad  \quad \quad  \quad \quad \quad  \quad\quad \quad  \quad \quad \quad  \quad \quad \quad (log scale - V/m)   \quad   \quad  \quad  \quad    (V/m)   \quad  \quad  \quad  \quad  \quad  \quad  \quad   (\%) \quad  \quad  \quad  \quad  \quad  \quad  \quad  \quad  \quad  \quad  \quad \quad  \quad  \quad \quad (rad) \quad \quad  \quad  \quad \quad  \quad \quad (rad)
\vspace{-12.0mm}
\end{minipage}
\raisebox{6ex}
{
    \begin{minipage}{0.1\linewidth}
        \raggedright 400 $\MHz$ 
    \end{minipage}
}
\includegraphics[width=0.47\textwidth, height=.11\textwidth]{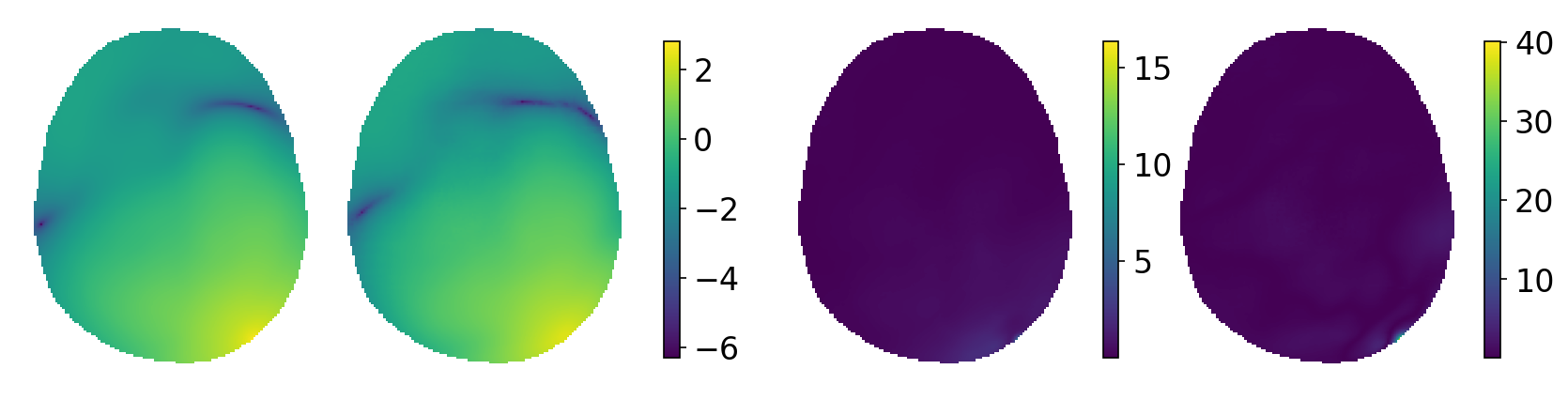}
\includegraphics[width=0.47\textwidth, height=.11\textwidth]{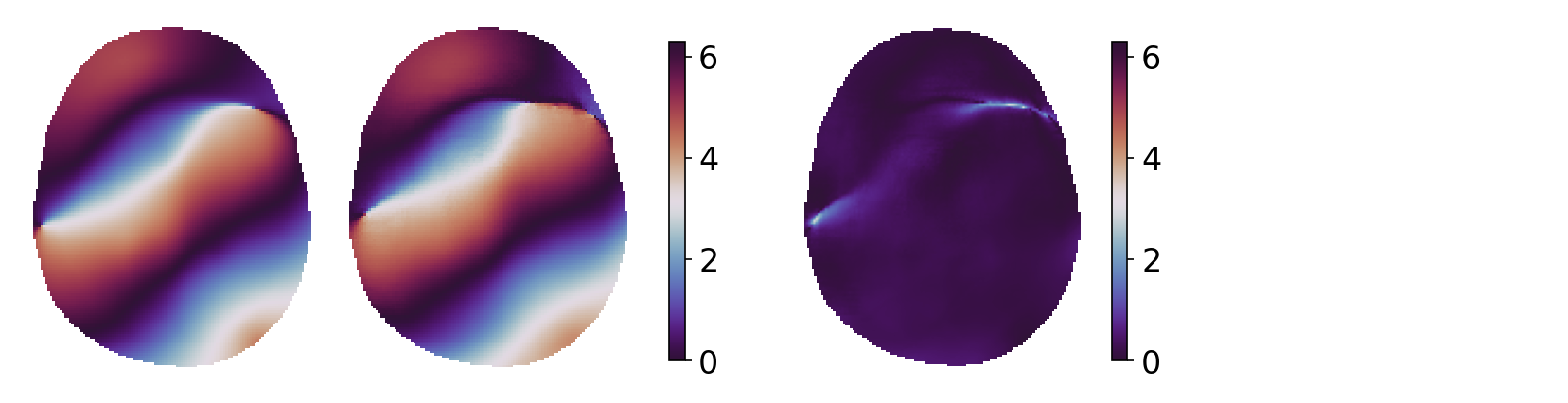}

\raisebox{6ex}
{
    \begin{minipage}{0.1\linewidth}
        \raggedright 900 $\MHz$ 
    \end{minipage}
}
\includegraphics[width=0.47\textwidth, height=.11\textwidth]{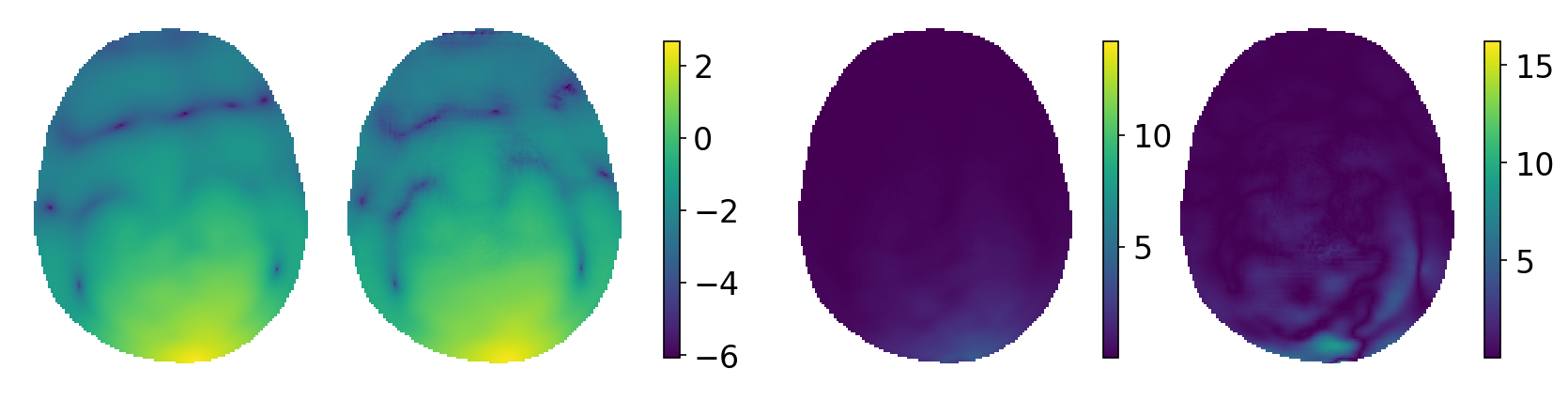}
\includegraphics[width=0.47\textwidth, height=.11\textwidth]{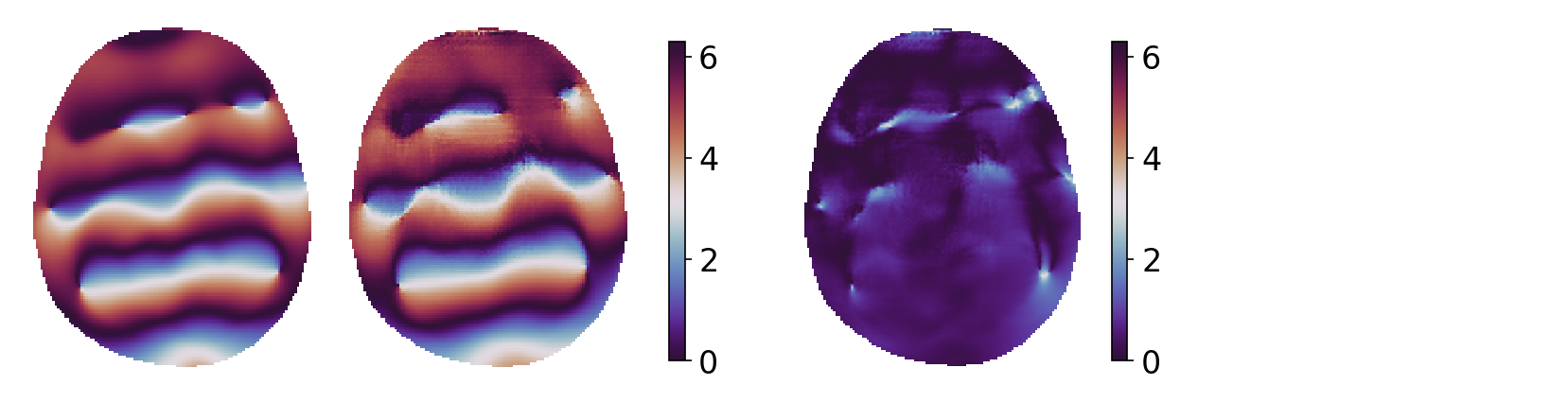}

\raisebox{6ex}
{
    \begin{minipage}{0.1\linewidth}
        \raggedright 1.5 $\GHz$ 
    \end{minipage}
}
\includegraphics[width=0.47\textwidth, height=.11\textwidth]{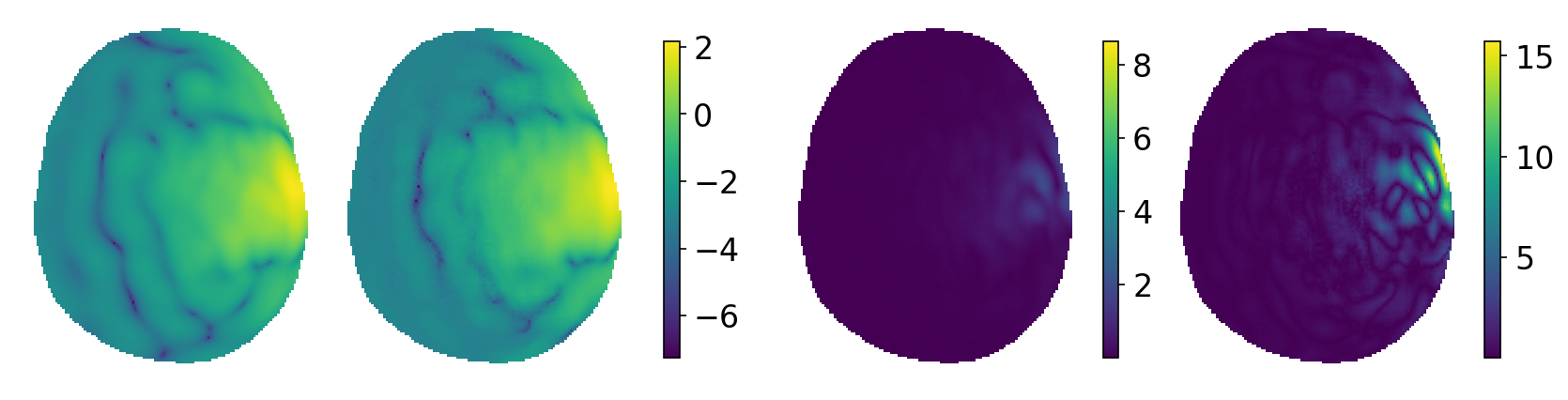}
\includegraphics[width=0.47\textwidth, height=.11\textwidth]{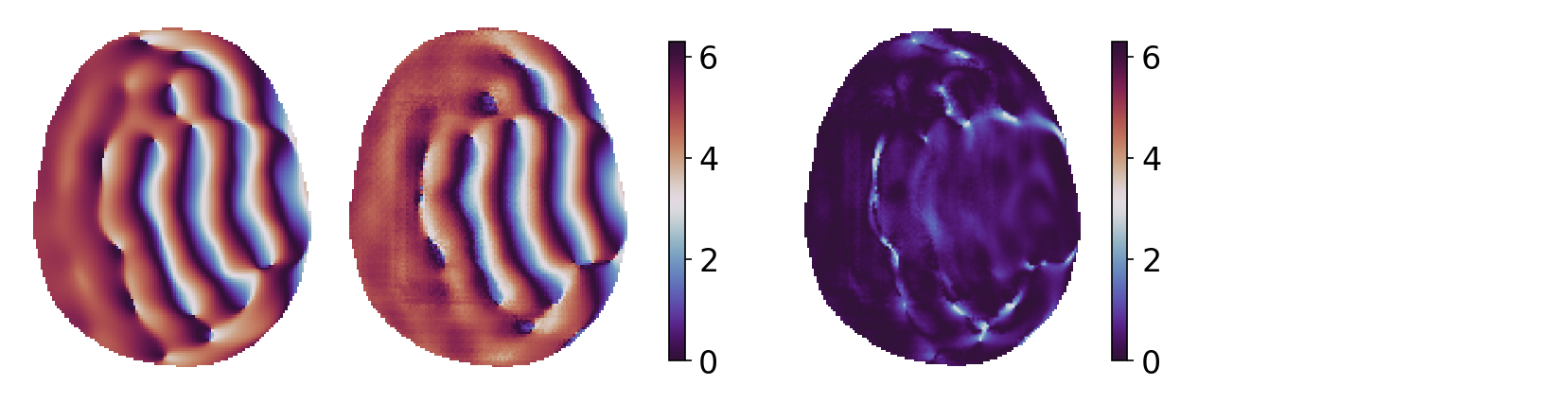}
\caption{Validation dataset evaluation of electric field: \textit{Top row.} At $400\MHz$, the amplitude of the electric field is plotted for ground truth $|y|$, model output $|\hat{y}|$,  AD (\ref{absdiff}), and RD (\ref{RelDiff}). Then, the  electric field phase is plotted for the corresponding label's ground truth angles $\angle y$, model output angles $\angle \hat{y}$, and PD (\ref{diff_eq_ang}). \textit{Middle and bottom rows.} Same as top row but for $900 \MHz$ and $1.5 \GHz$, respectively. Notice strong agreement between ground truth and model output across various frequencies.}\label{fig:400MHz_DT_MO_diff}
\end{figure*}

\subsection{U-Net model}\label{EMu:unetmodel}
 The U-Net architecture that forms the basis of EMulator, inspired from~\cite{ronneberger2015u}, is shown in Fig.~\ref{fig:pipeline}. The inputs consist of the human brain model with the antenna location incorporated. We specify the location of the antenna on the model image by assigning a high value to the pixel containing the antenna.  EMulator consists of four encoder-decoder stages. 
 

We denote $y \in \mathbb{C}^{158 \times 133}$   as the ground truth electric field response calculated by  COMSOL. The predicted electric field response $\hat{y} \in \mathbb{C}^{158 \times 133}$ is then calculated using EMulator. 
EMulator is trained to minimize the mean square error $(MSE) =  \frac{1}{N \times V}  \sum_{i=1}^{N}  \sum_{j=1}^{V}  |y_{ij} - \hat{y}_{ij}|^2$. Here,   $N = \{N_{train},  N_{val}\} = \{602, 258\}$ are the number of training and validation samples, respectively, and $V = 158 \times 133$ is the number of voxels in each training or validation sample.

\subsection{Evaluation metrics}\label{EMu:eva_metrics}

\subsubsection{Ground truth, model output, and difference plots}
We plot the ground truth $|y|$, model output $|\hat{y}|$, and the absolute difference (AD) between ground truth and model output as 
\begin{equation}
  \text{AD} = |y - \hat{y}|. \label{absdiff}
\end{equation}
Additionally, we also evaluate the percentage relative difference (RD)~\cite{jia2022deeptdcs}, which evaluates the error between the ground truth and the model output with respect to a baseline reference level of the maximum ground truth electric field intensity in the sample, given by
\begin{equation}\label{RelDiff}
\begin{aligned}
 \text{RD}  =  \frac{||y_v| - |\hat{y}_v||}{\max_{v} |y_v| } \times 100,
 \text{where } v \in V.
\end{aligned}
\end{equation}    
For phases, the phase difference (PD) is defined as the minimum  distance between the ground truth and the estimated phase on a circle, given by 
\begin{equation}
  \text{PD} = \text{min}(|\angle y - \angle \hat{y}|, 2\pi - |\angle y - \angle \hat{y}|). \label{diff_eq_ang}
\end{equation}

\subsubsection{Magnitude of the complex correlation coefficient}

We evaluate the magnitude of the complex correlation coefficient (CC)~\cite{schreier2010statistical} to determine the similarity between the ground truth and the model output.


\section{Results}\label{results}

\subsection{Training and validation datasets}
We evaluate training and validation datasets for three frequencies: $400 \MHz, 900 \MHz,$ and $ 1.5 \GHz$. The EMulator network is implemented and optimized in PyTorch~\cite{paszke2017automatic}. The U-Net code is adapted from~\cite{Unet_code_PT}. The code is executed with a single NVIDIA GeForce RTX 3090 Ti GPU. For each frequency, out of the 20 segmented human head models available~\cite{BrainWeb, BrainWebsite}, 14 subjects were used for training, and 6 subjects were used for validation.

\subsection{Ground truth, model output, and difference plots}

Fig.~\ref{fig:400MHz_DT_MO_diff} shows the performance of the trained model on the validation dataset when the antennas are operating at the frequency of   $400\MHz$, $900\MHz$ and $1.5\GHz$. The first column shows the ground truth electric field $|y|$ obtained from COMSOL for a randomly selected validation label at each frequency. The second column indicates the output of the trained model $|\hat{y}|$, and the third column shows the absolute difference between them, given by (\ref{absdiff}).
Next, the relative difference (\%) is plotted in the fourth column, given by (\ref{RelDiff}). From these plots, we can notice strong agreement between the ground truth  $|y|$ and model output  $|\hat{y}|$. 

For the next three successive plots, the first column indicates the corresponding ground truth angle  $\angle y$ obtained from COMSOL for that label, the second column gives the angle of model output $\angle \hat{y}$, and the last column gives the difference between the angles, given by (\ref{diff_eq_ang}). Once again, we notice strong agreement between the ground truth phases  $\angle y$ and model output phases $\angle \hat{y}$. We also notice in phase plots that the error could be high at specific locations ($\sim 2\pi$). However, these errors occur at places where the amplitude of the electric field itself is very low (see the corresponding  amplitude label from the first three rows), potentially making this error insignificant, as the electric field value at those locations is significantly low.
The low absolute difference plots show strong agreement between ground truth and model output across randomly selected validation labels, demonstrating that EMulator can be used to calculate the electric field at trained antenna locations at various carrier frequencies with high accuracy.

\subsection{Per subject statistics}
\begin{figure}[htb!]
\centering
\includegraphics[width=0.45\textwidth, height=.16\textwidth]{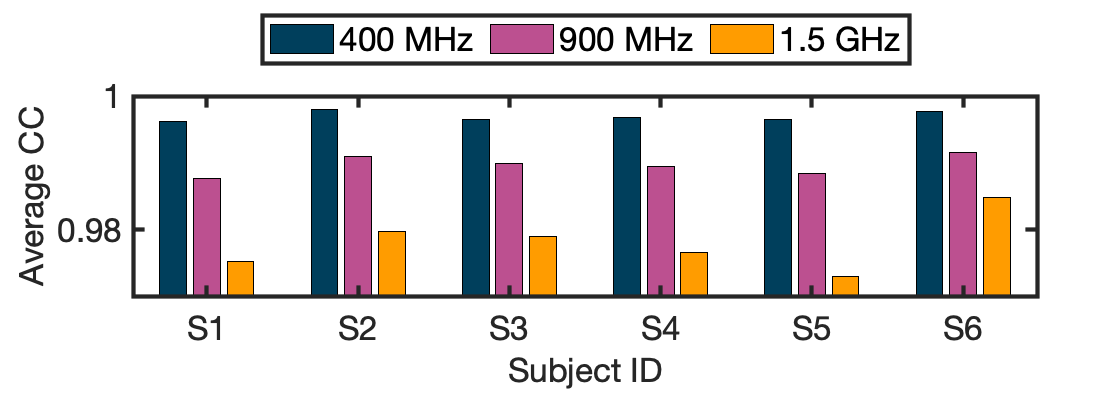}
\caption{Per subject statistics for the 6 validation subjects show a high average CC of more than 0.97 across all the frequencies.}
\label{fig:pp_stats}
\end{figure}
The CC for all the 6 validation subjects is plotted in Fig.~\ref{fig:pp_stats}. First, we see a strong average correlation of more than 0.97 across all the subjects and across all the frequencies. Despite strong correlation, notice that our results show that it is harder to estimate higher frequencies.
There are two reasons for this: first,  the higher frequencies have a faster variation in wavenumber that needs to be captured by EMulator. Secondly, due to higher attenuation, higher frequencies have a larger range between minimum and maximum values, making it computationally harder for the EMulator to learn the salient features.


\subsection{Computation time: COMSOL versus EMulator}

\begin{figure}[htb!]
\centering
\includegraphics[width=0.5\textwidth, height=.35\textwidth]{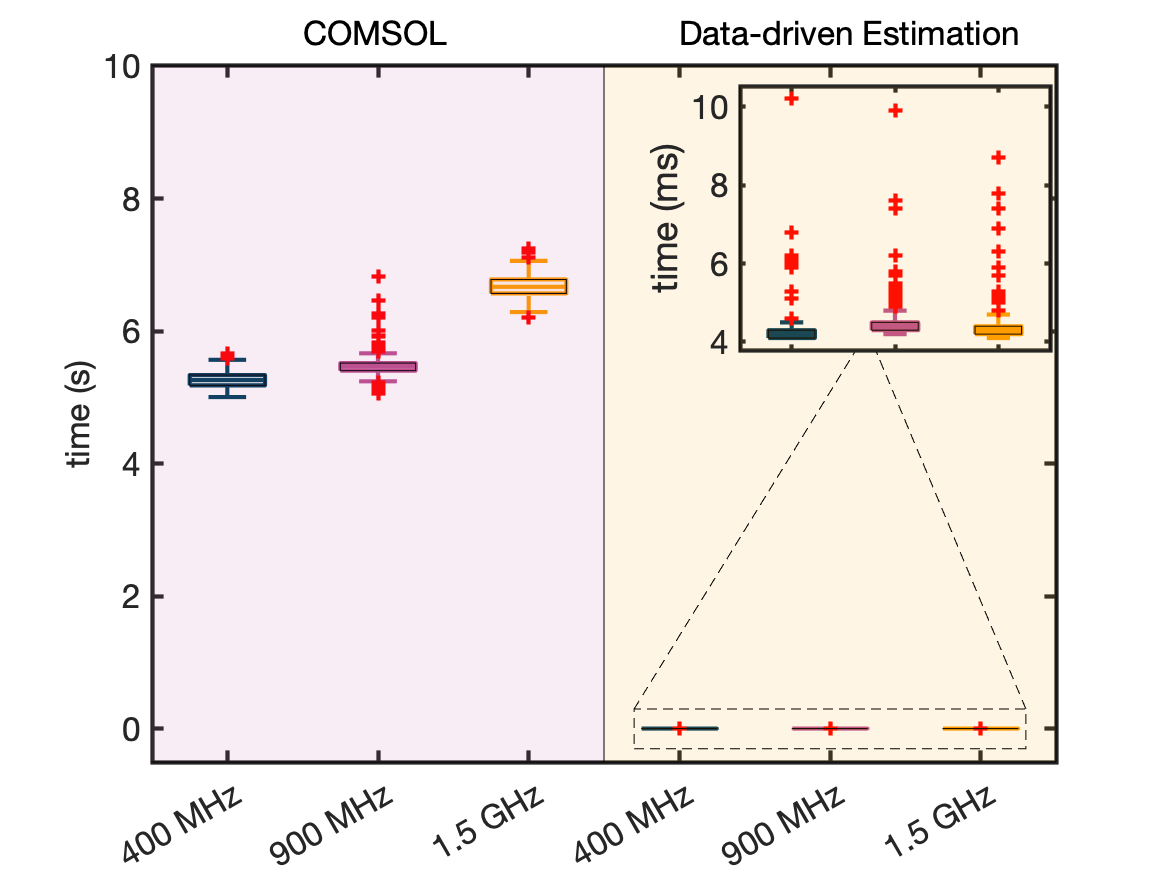}
\caption{Time taken by COMSOL versus EMulator to perform electric field estimation across the three frequencies. Data-driven  estimation is on average more than $\times 1200$ faster than COMSOL.}
\label{fig:t_comsol}
\end{figure}

Fig.~\ref{fig:t_comsol} shows the time taken for electric field estimation for each of the 6 validation subjects across the three frequencies for COMSOL and EMulator. We notice that the higher the frequency, the more time  COMSOL takes to perform electric field estimation. This is because higher frequencies need finer meshes to  compute the electric field accurately. Whereas, the time taken by EMulator to compute the electric field remains the same. That's because the computational time taken by EMulator depends upon the model size, and, in our case, the model size is the same for all  three frequencies. Another key takeaway is that, on average, data-driven  electric field estimation is more than $\times 1200$ faster than COMSOL electric field estimation.

\subsection{Interpolation capability test}
\begin{figure}[htb!]
\centering
\includegraphics[width=.45\textwidth, height=.17\textwidth]{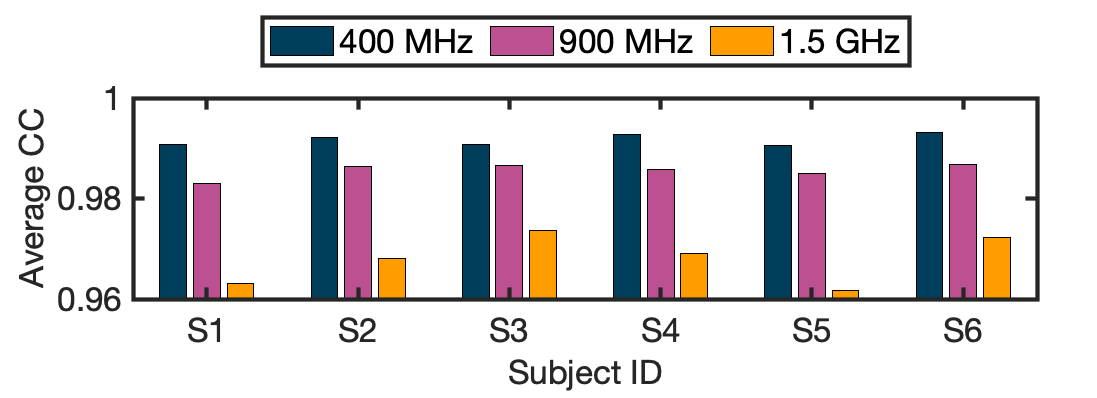}
\caption{Per subject statistics for the 6 validation subjects on the non-trained antenna locations. This captures the interpolation capability of EMulator.}
\label{fig:20_interp_antz_stats}
\end{figure}

We also tested the interpolation capability of our trained model for non-trained antenna locations. For this purpose, we place antennas at 20 random locations around the brain tissue of the validation subjects and calculate ground truth electric fields from COMSOL. We then compute the electric field generated by our model for these non-trained antenna locations. Fig.~\ref{fig:20_interp_antz_stats} shows high CC between the ground truth and the estimated electric field across all the 6 subjects for non-trained antenna locations. Notice that a high average correlation of  0.96 and above is maintained across all the frequencies. This shows that, using our model, we can predict electric field even for non-trained antenna locations without running the computationally expensive FEM software COMSOL, drastically speeding up the optimizations.

\section{Discussion \& Conclusion}\label{discuss}
In this work, we developed a data-driven method of estimating the electric field generated by antenna arrays inside the brain tissue in real-time. We trained EMulator, a U-Net inspired architecture, to achieve low training and validation error from the ground truth electric field generated by COMSOL. Our work showed high average correlation coefficient of more than $0.97$ across all the validation subjects and frequencies for the trained antenna locations. 
It also showed a strong correlation of more than $0.96$ for non-trained antenna locations for the validation dataset.
On average, our electric field estimation is approximately $\times 1200$ faster than COMSOL. For future research, this opens interesting avenues for fast 3D data-driven electric field modeling for clinical applications. We will also employ synthetic data augmentation techniques to increase the robustness of EMulator.
Our method is generic and can be utilized to improve the targeting capability of other forms of neuromodulation, which optimize the locations of one or more transducers~\cite{TES_strength, sorkhabi2020temporally, liu2022noninvasive}. Under a  broader scope, since  our model also demonstrated robust interpolation capability, problems in operations research that have mixed-integer non-linear problems could also benefit from our approach of making surrogate models that can interpolate to give results at non-integer locations. 


\bibliographystyle{ieeetr}
\bibliography{IEEEexample}

\end{document}